\documentclass[bbm,amsfonts,amsmath,amssymb,12pt]{article}
\usepackage{epsfig}
\usepackage{graphicx}
\usepackage{dcolumn}
\usepackage{bm}
\usepackage{color}
\usepackage{amsmath}

\definecolor{red}{rgb}{1., 0., 0.}
\definecolor{blue}{rgb}{0., 0., 1.}
\definecolor{green}{rgb}{0.1, 0.7, 0.}
\definecolor{purp}{rgb}{1,0,1}

\def\drawbox#1#2{\hrule height#2pt
        \hbox{\vrule width#2pt height#1pt \kern#1pt
              \vrule width#2pt}
              \hrule height#2pt}

\def\Fund#1#2{\vcenter{\vbox{\drawbox{#1}{#2}}}}
\def\Asym#1#2{\vcenter{\vbox{\drawbox{#1}{#2}
              \kern-#2pt       
              \drawbox{#1}{#2}}}}

\def\fund{\Fund{6.4}{0.3}}

\newcommand {\beq} {\begin{equation}}
\newcommand {\eeq} {\end{equation}}
 \newcommand{\be}{\begin{eqnarray}}
\newcommand{\ee}{\end{eqnarray}}

\begin{document}

\begin{titlepage}


\vspace {1cm}

\centerline{{\Large \bf Extending the Veneziano-Yankielowicz
Effective Theory}}

\vskip 1cm \centerline{\large P. Merlatti and F. Sannino} \vskip
0.1cm

\vskip 0.5cm \centerline{NORDITA, 
Blegdamsvej 17,
\, DK-2100 Copenhagen \O, Denmark }

\vskip 1cm

\begin{abstract}
We extend the Veneziano Yankielowicz (VY) effective theory in
order to account for ordinary glueball states. We propose a new
form of the superpotential including a chiral superfield for the
glueball degrees of freedom. When integrating it ``out'' we obtain
the VY superpotential while the N vacua of the theory naturally
emerge. This fact has a counterpart in the Dijkgraaf and Vafa
geometric approach. We suggest a link of the new field with the
underlying degrees of freedom which allows us to integrate it
``in'' the VY theory. We finally break supersymmetry by adding a
gluino mass and show that the K\"{a}hler independent part of the
``potential'' has the same form of the ordinary Yang-Mills
glueball effective potential.
\end{abstract}

\end{titlepage}

\section{Introduction}
\label{uno} Supersymmetric gauge theories are much studied in the
hope that one day they may be relevant to understand the physics
of the real world. We already know a great deal about
supersymmetric gauge theories which are closer to their non
supersymmetric cousins, namely ${\cal N}=1$ supersymmetric gauge
theories, see \cite{Intriligator:1995au} for a review.

Effective Lagrangians are an important tool for describing
strongly interacting theories in terms of their relevant degrees
of freedom. A well known effective Lagrangian which economically
describes the vacuum structure of super Yang-Mills has been
constructed by Veneziano and Yankielowicz (VY)
\cite{Veneziano:1982ah}. The Lagrangian concisely summarizes the
symmetry of the underlying theory in terms of a ``minimal'' number
of degrees of freedom which are encoded in the superfield $S$
\begin{eqnarray}
S=\frac{3}{32\pi^2\, N}{\rm Tr}W^2 \ ,
\end{eqnarray}
where $W_{\alpha}$ is the supersymmetric field strength. When
interpreting $S$ as an elementary field it describes a gluinoball
and its associated fermionic partner. In this paper we follow the
notation introduced in \cite{Sannino:2003xe}.

Besides the gluinoballs with non zero $R$-charge also glueball
states with zero $R$ charge are {\it important} degrees of
freedom. These states are expected to play a relevant role when
breaking supersymmetry by adding a gluino mass term. This is so
since the basic degrees of freedom of the pure Yang-Mills theory
are glueballs. {}Further support for the relevance of such
glueball states in super Yang-Mills comes from lattice simulations
\cite{Feo:2002yi}. Recently it has also been argued that certain
non supersymmetric theories, named orientifold, at infinite number
of colors share a number of properties of the ordinary super
Yang-Mills theory \cite{Armoni:2004uu}. With such theories one
can, in principle, interpolate \cite{Armoni:2004uu} between super
Yang Mills and QCD with one dirac flavor. One can also imagine a
different large $N$ limit \cite{Corrigan:1979xf,Armoni:2004uu}. We
were also able to obtain a number of relevant results by including
the leading $1/N$ corrections via an effective Lagrangian approach
\cite{Sannino:2003xe}. Since glueballs are present in QCD, and for
a generic orientifold theory, we expect their presence also at
large $N$, i.e. in the super Yang-Mills limit. It is, hence, very
natural to expect these states to be present at low energies in
super Yang-Mills.

However no {\it physical} glueballs appear in the VY effective
Lagrangian. In this paper we extend the VY Lagrangian to take into
account the glueball states. Some attempts have already appeared
in the literature
\cite{Shore:1982kh,Kaymakcalan:1983jh,Farrar:1997fn,
Farrar:1998rm,Gabadadze:1998bi,Bergamin:2003ub,Sannino:1997dd,Cerdeno:2003us}.
{}Shore \cite{Shore:1982kh} as well as Kaymakcalan and Schechter
\cite{Kaymakcalan:1983jh} proposed to use controgradient fields to
include glueball states in the theory. However due to the
classical field constraints used in this approach supersymmetry
was not guaranteed to hold at the effective Lagrangian level.
Another approach \cite{Farrar:1997fn,Cerdeno:2003us} has been to
rewrite $S$ as the field strength of a real gauge superfield
associated to a $3$-form. In order to introduce the glueball
fields here a model dependent breaking of the gauge invariance has
been used.

In this paper we introduce immediately a chiral superfield $\chi$
with the quantum numbers of a glueball. The basic constraints
which we will use to construct the effective superpontetial
involving $S$ and $\chi$ are: i) The superpotential reproduces the
anomalies of super Yang-Mills; ii) The vacuum structure is
unaltered even in the presence of the glueball field. These two
requirements lead to a general form of the superpotential in terms
of an undetermined function of the chiral field $f(\chi)$. However
we will argue in favor of a specific form for $f(\chi)$ which has
a number of amusing properties. {}For example the $N$ vacua of the
theory emerge naturally when integrating out the glueball
superfield $\chi$. This intriguing relation has also a counterpart
in the geometric approach to the effective Lagrangian theory
proposed by Dijkgraaf and Vafa \cite{Dijkgraaf:2002dh} or in the
more field-theoretical approach presented in
\cite{Cachazo:2002ry}. We have also suggested an integrating in
procedure which surprisingly yields exactly the specific form of
the function $f(\chi)$.

 Another important check is associated to
supersymmetry breaking. When adding a gluino mass to the theory
\'{a} la Masiero and Veneziano \cite{Masiero-Veneziano} the same
choice of the function $f(\chi)$ leads to a K\"{a}hler independent
part of the ``potential'' which has the same functional form of
the glueball effective potential for the non supersymmetric
Yang-Mills theory developed and used in
\cite{schechter,joe,MS,SST}.

We provide the link with the underlying degrees of freedom. This
is done by first providing a classical relation between the
glueball superfield $\chi$ and the contragradient fields built out
of $S$ which mimics the one employed by Shore \cite{Shore:1982kh}
as well as Kaymakcalan and Schechter \cite{Kaymakcalan:1983jh} and
then upgrading the classical constraint to a quantum one. Due to
the nature of the quantum constraint the effective theory in
general preserves supersymmetry. We also briefly review the
three-form approach while outlining a possible way of linking the
two approaches.

 In section \ref{due} we briefly review the VY theory and set the
notation. Section \ref{tre} is devoted to the extension of the VY
theory and contains a number of subsections in which we provide
consistency checks for the proposed extension of the VY
superpotential. Finally we conclude in section \ref{quattro}.

\section{Reviewing the VY effective Lagrangian}
\label{due}
 It is instructive to briefly review the VY Lagrangian while
 introducing the notation as in \cite{Sannino:2003xe}.

The underlying Lagrangian of SU($N$) supersymmetric gluodynamics
is\,\footnote{The Grassmann integration is defined in such a way
that $\int \, \theta^2\, d^2\theta =1$.} \be {\cal L}&
=&\frac{1}{2\,g^2}\,
\int \! {\rm d}^2 \theta\, \mbox{Tr}\,W^2 +{\rm H.c.} \nonumber\\[3mm]
& =&-\frac{1}{4 g^2} \,  G_{\mu\nu}^a G^{a\mu\nu}
+\frac{1}{2g^2}\, D^a D^a +\frac{i}{g^2} \lambda^a\sigma^\mu {\cal
D}_\mu\bar \lambda^a \, , \ee where $g$ is the gauge coupling, the
vacuum angle is set to zero and \beq \mbox{Tr}\,W^2 \equiv
\frac{1}{2}W^{a,\alpha}W^{a}_{\alpha}= -
\frac{1}{2}\lambda^{a,\alpha}\lambda^{a}_{\alpha}\,. \eeq  The low
energy effective superpotential \cite{Veneziano:1982ah}
constructed in terms of the composite chiral superfield $S$,
\beq S= \frac{3}{32\pi^2 N }\,\mbox{Tr}\,W^2 \,, \eeq
is:
 \be {W}_{VY}&=&
\frac{2N}{3} \int \! {\rm d}^2\theta\, \left\{ S  \ln
\left(\frac{S }{\Lambda^3}\right)^{N}-NS\right\}  \, , \label{VY}
\nonumber \\\ee
where $\Lambda$ is a renormalization group invariant scale.

The chiral superfield $S$ at the component level has the standard
decomposition $S(y)=\varphi(y) + \sqrt{2} \theta \chi(y) +
\theta^2 F(y)$, where $y^\mu$ is the chiral coordinate,
$y^\mu=x^\mu - i \theta \sigma^\mu \bar{\theta}$, and
\begin{eqnarray}
\varphi &=& \frac{3}{64\pi^2 N}\left[-\lambda^{a,\alpha}\lambda^{a}_{\alpha}\right] \ ,\\
\sqrt{2}\chi &=& \frac{3}{64\pi^2 N}
\left[G^a_{\alpha\beta}\lambda^{a,\beta} +2i D^a \lambda^{a}_{\alpha} \right]\ , \\
F &= &\frac{3}{64\pi^2 N}\left[-\frac{1}{2} G^a_{\mu\nu}
G^{a\mu\nu}+ \frac{i}{2}G^a_{\mu\nu}
\tilde{G}^{a\mu\nu}+\mbox{f.t.} \right] \ , \label{decomp}
\end{eqnarray}
where f.t. stands for fermion terms.

The complex field $\varphi$ represents the scalar and pseudoscalar
gluino-balls while $\chi$ is their fermionic partner. Although it
is tempting to say that $F$ represents the scalar and the
pseudoscalar glueball it is an auxiliary field. Hence these states
are not represented in the VY Lagrangian.

\section{Introducing the Glueball Superfield $\chi$}
\label{tre} One of the hardest problems in confining theories is
the identification of the relevant degrees of freedom at low
energies especially when the latter are not phenomenologically
known. If introduced it is even harder to find their relation with
the underlying gauge theory. What is straightforward though is the
quantum number identification of the states of interest.

Here, assuming that the lowest component of the chiral superfield
$\chi$ contains a scalar glueball, we deduce that the chiral
superfield has $R$-charge zero. Covariance under superconformal
transformations leads to the relation:
\begin{equation}\label{charge}
d=\frac{3}{2}\, n_R \ ,
\end{equation}
between the mass dimension $d$ of a generic chiral superfield and
its $R$-charge $n_R$\footnote{Actually a generic field
transforming properly (i.e. covariantly) under the superconformal
group must satisfy the constraint \cite{Shore:1982kh}
$2n~=~d~=~\frac{3}{2}n_R$,
where $n$ is the charge of the field under the superconformal
transformations. One can show that invariance under dilatations
implies invariance under superconformal transformations.}. The
glueball superfield $\chi$ has then zero mass dimension.

Due to the $\chi$ properties just found the general effective
superpotential saturating all of the relevant anomalies and
containing both $S$ and $\chi$ is:
\begin{eqnarray}
W\left[S,\chi \right]\, = \,\frac{2N}{3}
\left\{ S  \ln \left(\frac{S }{\Lambda^3}\right)^{N}-NS   -
S\,f(\chi) \right\}
\label{newsuper}
\end{eqnarray}
with $f(\chi)$ an holomorphic function of $\chi$. The latter in
components reads:
\begin{eqnarray}
\chi=\varphi_{\chi} + \sqrt{2}\theta \psi_{\chi}  + \theta^2
F_{\chi} \ .
\end{eqnarray}
The determination of the function $f(\chi)$ would shed light on
the super Yang-Mills infrared properties. We will provide various
arguments pointing to the function:
\begin{eqnarray}\label{effe}
f(\chi) = N\ln \left[-e\frac{\chi}{N}\ln \chi^N\right] \ .
\end{eqnarray}
This function passes a number of consistency checks: i) We recover
the VY superpotential when the glueball superfield is integrated
out. Besides this procedure naturally leads to the $N$ independent
vacua of the theory. ii) We can now better approach non
supersymmetric gluondynamics when giving a mass to the gluino. The
theory leads to a potential which resembles the ordinary glueball
effective potential for the Yang-Mills theory. iii) The
superpotential in eq.~(\ref{newsuper}) has a natural
interpretation in the geometric approach to the effective
Lagrangian theory proposed by Dijkgraaf and Vafa. iv) A reasonable
integrating in method leads to the same function.

\subsection{Integrating Out $\chi$}

The first non trivial check comes from integrating out the field
$\chi$ via its equation of motion:
\begin{eqnarray}
\frac{\partial W [S,\chi]}{\partial \chi}&=&
-\frac{2N^2}{3}\,\frac{S}{\chi\,\ln \chi^N}\left[\ln \chi^N +
N\right] =  \nonumber \\ &=& -\frac{2N^2}{3}\,\frac{S}{\chi\,\ln
\chi^N}\left[\ln_{(0)} \chi^N + 2\pi\,i\,k + N\right] = 0 \ ,
\end{eqnarray}
where we have made explicit the dependence of the logarithm on the
branches. $\ln_{(0)} \chi^N$ is by definition the $k=0$ branch of
$\ln \chi^N$, i.e. its imaginary part lies within $0$ and $2\pi$.
The solution is:
\begin{eqnarray}
\chi=\frac{1}{e} \, e^{-2\pi\,i\frac{k}{N}} \ ,
\end{eqnarray}
which yields:
\begin{eqnarray}\label{outchi}
W_k\left[S\right]\, = \,\frac{2N}{3}
\left\{ S  \ln \left(\frac{S }{\Lambda^3
e^{-2\pi\,i\frac{k}{N}}}\right)^{N}-NS\right\} \ .
\end{eqnarray}
This reproduces the standard VY result. Besides in this way we can
also account naturally for the $N$ vacua of super Yang-Mills.
Somewhat surprisingly the present superpotential is equivalent to
the one proposed by Kovner and Shifman in \cite{Kovner-Shifman}.
The summation over the $k$ branches is now automatic since from
the start we needed to integrate over all of the allowed field
$\chi$ configurations in the path integral. After having
eliminated the field $\chi$ we still have to sum over the $k$
branches of the logarithm.

\subsection{Approaching the Yang-Mills Theory}

An important test for the proposed generalization of the VY
superpotential deals with the effects of a gluino mass term and
the Yang-Mills limit.

The most straightforward approach is to add a ``soft''
supersymmetry breaking term to the Lagrangian. Masiero and
Veneziano \cite{Masiero-Veneziano} introduced the following gluino
mass term,
\begin{eqnarray}
\Delta {\cal L}_{m} = -\frac{m}{2g^2} \, \lambda \lambda + {\rm
h.c.} \ .
\end{eqnarray}
which at the effective-Lagrangian level translates as
\begin{equation}
\Delta {\cal L}_{m}=\frac{m}{g^2} \frac{N}{3}\,32
\pi^2\,\left(\varphi + \bar{\varphi}\right) = \frac{4\,
m}{3\lambda}\, N^2 \left(\varphi + \bar{\varphi}\right)\,,
\label{soft}
\end{equation}
where we introduced the 't Hooft coupling \beq \lambda\equiv
\frac{g^2 N}{8\pi^2}\,. \eeq It is convenient to assume the mass
parameter $m$ to be real and positive. One can always make it real
and positive by redefining the vacuum angle $\theta$. In what
follows we will adopt this convention.

The softness restriction is $m /\lambda  \ll {\Lambda}$. Recently,
soft SUSY breaking has been reanalyzed in \cite{EHS}, while a
model for not-soft breaking has been proposed in
\cite{Sannino:1997dd}.

Note that the combination $m/\lambda$ is renormalization-group
invariant to leading order, and scales as $N^0$; the one which is
renormalization-group invariant to all orders can be found too,
see \cite{Hisano:1997ua}.  Analysis
 of this model indicates that the theory is
``trying'' to approach the non-SUSY Yang-Mills case. Namely, the
spin-$0$ and spin-$1/2$ particles split from each other, and their
masses each pick up a piece linear in $m$. One of the $N$ distinct
vacua of the SUSY theory becomes the true minimum. Furthermore,
the vacuum value of the gluon  condensate is no longer zero.

Although the previous results were encouraging, the glueball
states were yet not accounted for in the standard VY approach
followed in \cite{Masiero-Veneziano}. Hence it is hard to imagine
how the Yang-Mills theory may have emerged after supersymmetry
breaking.

In our case we have glueball states even in the supersymmetric
limit. We expect, by providing a mass to the gluino field, that
these states go over the ordinary Yang-Mills glueball fields. To
better illustrate this phenomenon we focus on the part of the
potential of the theory which is K\"{a}hler independent. At the
first order in the gluino mass, after having integrated out the
gluino field, one derives:
\begin{eqnarray}
{V[\varphi_{\chi}]}_k = \frac{4\, m \Lambda^3}{3\lambda}\, N^2 \,e
\left[\varphi_{\chi}\ln_{(0)}\varphi_{\chi}  +
\bar{\varphi}_{\chi} \ln_{(0)} \bar{\varphi}_{\chi} +
2\pi\,i\,\frac{k}{N}\left(\varphi_{\chi} - \bar{\varphi}_{\chi}
\right)\right] + \cdots \ . \label{soft}
\end{eqnarray}
The dots indicate the K\"{a}hler dependent terms which do not
contribute to the vacuum expectation value of the potential, do
not affect the vev of $\varphi_{\chi}$ at leading order in $m$ and
will be dropped in the following. $\varphi_{\chi}$ is the
dimensionless glueball field. The ground state is then obtained
for:
\begin{eqnarray}
e\,\varphi_{\chi}\, = \, e^{-2\pi\,i\,\frac{k}{N}} \ ,
\end{eqnarray}
and the potential is minimized for $k=0$
\begin{eqnarray}
\langle V \rangle = -8\frac{m\Lambda^3}{3\lambda}\, N^2\, {\rm
min}_{k} \left\{ \cos 2\pi\,\frac{k}{N} \right\} \qquad
k=0,1,\ldots N-1 .
\end{eqnarray}
In this way the degeneracy of the $N$ supersymmetric vacua is
lifted and only the $k=0$ solution is selected as the ground state
of the theory.

Due to the presence of the single logarithm term this potential
resembles the effective potential for ordinary pure Yang-Mills
theory \cite{schechter,joe,MS,SST}. In order to make this
similarity even more transparent, and working in the $k=0$ branch,
we define the field $F_{Y} = \frac{8\, m \Lambda^3}{3\lambda} \,e
\varphi_{\chi}$ which has mass dimension four. The potential
becomes:
\begin{eqnarray} V[F_{Y}] = \frac{N^2}{2}\left[F_{Y} \ln
\frac{F_{Y}}{\Lambda_{Y}^4} + {\rm c.c.}\right] \ ,
 \label{vfinale}\end{eqnarray}
with $\Lambda_{Y}^4 =  8\,m \,\Lambda^3\,e/(3\lambda) $. The
parameter $\Lambda_{Y}$ is not the pure Yang-Mills scale yet. We
recall that one loop decoupling gives the following relation
between the super Yang-Mills scale, the gluino mass and the
Yang-Mills scale $\Lambda_{YM}=m^{2/11}\Lambda^{9/11}$. All the
scales become comparable, i.e. $\Lambda_{Y}\sim \Lambda_{YM}\sim
\Lambda$ in the limit $m\sim \Lambda$. {}By computing, using
eq.~(\ref{vfinale}), the trace of the energy momentum tensor we
obtain:
\begin{eqnarray}
{\theta_Y}^{\mu}_{\mu} = 4\,V-4\,\left[F_{Y}\frac{\partial
V}{\partial F_{Y}} + {\rm c. c}\right] =
-2\,{N^2}\left[F_{Y}+\bar{F}_{Y}\right] \ .
\end{eqnarray}
If we now imagine the gluino already decoupled at a scale $m \sim
\Lambda$ we can identify this expression with the one associated
with the trace of the energy momentum tensor of the underlying
pure Yang-Mills theory which is
\begin{eqnarray}
{\theta_{YM}}^{\mu}_{\mu} =
-\frac{11\,N}{3}\frac{1}{32\pi^2}G^{\mu\nu}G_{\mu\nu} \ .
\end{eqnarray}
One then finds that $Re(F_{Y}) \propto G^a_{\mu\nu} G^{a\mu\nu}$
while it is natural to expect that $F_{Y} \propto G^a_{\mu\nu}
G^{a\mu\nu} - {i}\,G^a_{\mu\nu} \tilde{G}^{a\mu\nu}$. This also
supports the toy model approach used in \cite{Sannino:1997dd}.

We note that when supersymmetry remains intact the lowest
component of the superfield $\chi$ cannot be simply $G^a_{\mu\nu}
G^{a\mu\nu} - {i}\,G^a_{\mu\nu} \tilde{G}^{a\mu\nu}$.

Another amusing property is that the imaginary field of $F_{Y}$ is
unstable in the previous potential (see \cite{Sannino:1997dd} for
a modern discussion). This corresponds to a negative mass square
for the pseudoscalar glueball. Such a property has been a key
point for solving the $U(1)_A$ problem at the effective Lagrangian
level when quarks were added to the theory.

\subsection{Integrating In $\chi$}

After the interlude on supersymmetry breaking and the approach to
the Yang-Mills theory we now apply the integrating in method for
$S$ and $\chi$. In order to do so we first observe that the super
Yang-Mills superpotential evaluated on the vacuum of the theory
is:
\begin{eqnarray}
W_{\rm vac}\left[\Lambda,k \right]=-\frac{2}{3}N^2 \Lambda^3
e^{-2\pi\,i\frac{k}{N}}=-\frac{2}{3}N^2 \mu^3 \,e^{2\pi i\,
\frac{\tau - k}{N}}\ .
\end{eqnarray}
It is immediate to see the parameter $k$ as a shift of the super
Yang-Mills complexified coupling constant $\tau$:
\begin{eqnarray}
\tau=\frac{\theta}{2\pi} + i\,\frac{4\pi}{g^2} \ .
\end{eqnarray}
$S$ can be integrated in as explained in some detail in
\cite{Intriligator:1995au}. In this case the source for $S$ is
$\ln\Lambda^{3N}$. This logarithm term is generated at the one
loop level in super Yang-Mills.
We will return to the integrating in procedure for
$S$ at the end of this paragraph.

Here we simply upgrade the discrete shift of the complexified
coupling constant to a source for the c-number field $\chi$ and
define the superpotential linear in the source:
\begin{eqnarray}
W_{\rm linear}\left[\Lambda,D,\chi\right] =
-\frac{2N^2}{3}\Lambda^3\,D\,\chi \ , \quad {\rm with} \quad D
=-2\pi\,i\frac{k}{N} \ .
\end{eqnarray}
We will provide a justification of this term in the next section
when we will also discuss the relation between $\chi$ and the
underlying degrees of freedom of the theory.

The relation between $D$ and the vev of $\chi $ is obtained via:
\begin{eqnarray}
\frac{\partial W_{\rm vac}\left[\Lambda,D\right]}{\partial D} =
-\frac{2N^2}{3} \Lambda^3\chi \ ,
\end{eqnarray}
which yields:
\begin{eqnarray}
 D = \ln_{(0)}\chi \quad {\rm or~equivalently} \quad \ln_{(k)}\chi^N\equiv\ln_{(0)}\chi^N +
2\pi\,i\,k=0 \ , \label{relazione}
\end{eqnarray}
defining the branch $k$ of $\ln\chi^N$. The dynamical
superpotential as function of $\chi$ is then
\cite{Intriligator:1995au}:
\begin{eqnarray}
W\left[\Lambda,\chi\right] &=& \frac{2N^2}{3}\Lambda^3\,
\chi\left[2\pi i\,\frac{k}{N}\right]+ W_{\rm
vac}\left[\Lambda,D\right ] - W_{\rm
linear}\left[\Lambda,D,\chi\right] \nonumber \\& =&\frac{2N^2}{3}
\Lambda^3 \,\frac{\chi}{N}\, \left[ \ln_{(0)}
\left(\frac{\chi}{e}\right)^N + 2\pi i\,{k} \right]\equiv
\frac{2N^2}{3} \Lambda^3 \left[\frac{\chi}{N} \ln_{(k)}
\left(\frac{\chi}{e}\right)^N\right] \ , \nonumber \\
\end{eqnarray}
where we have substituted the relation between $D$ and $\chi$
given in eq.~(\ref{relazione}).

{}We now determine the complete superpotential by finally
integrating in $S$. We recall the aforementioned one loop relation
between $\Lambda$ and $S$ which in our normalization reads:
\begin{eqnarray}
W_{\rm loop}\left[\Lambda,S\right] = -\frac{2}{3}N S \,\left[\ln
\frac{\Lambda^{3N}}{\Lambda_0^{3N}}\right] \ ,
\end{eqnarray}
where $\Lambda_0$ is a reference scale and $\Lambda$ is the the
renormalization invariant scale of the theory. This requires:
\begin{eqnarray}
\frac{\partial W\left[\Lambda,\chi\right]}{\partial \ln \Lambda^3}
= -\frac{2}{3}N^2 S\ . \end{eqnarray} After performing the
innocuous field redefinition $\chi \rightarrow e\chi$ this leads
to
\begin{eqnarray}  S =
\Lambda^3 \left[-e\frac{\chi}{N} \ln\chi^N\right] \ ,
\label{relazione2}
\end{eqnarray}
where $S$ and $\chi$ are understood as vacuum expectation values
and we have dropped the subscript $(k)$ denoting the branch of the
logarithm. The complete effective superpotential as function of
$S$ and $\chi$ is:
\begin{eqnarray}
W\left[S,\chi\right] &=&  -\frac{2}{3}N S \,\left[\ln
\frac{\Lambda^{3N}}{\Lambda_0^{3N}}\right] +
W\left[\Lambda(S,\chi),\chi\right] - W_{\rm loop}
\left[\Lambda(S,\chi),S\right] \nonumber \\&& \nonumber
\\&=& \frac{2N}{3}S\left[\ln\left(\frac{S}{\Lambda^3}\right)^N -N
-N\ln \left(-e\frac{\chi}{N} \ln\chi^N\right) \right] \ ,
\end{eqnarray}
where for $\Lambda(S,\chi)$ we have used the relation in
eq.~(\ref{relazione2}). We have derived the desired function
$f(\chi)$.

\subsection{Relation with the fundamental degrees of freedom}

The effective Lagrangian describes two independent chiral
superfields, i.e. $S$ and $\chi$. While for $S$ we have an
interpretation in terms of the fundamental fields for $\chi$ we
still lack such an identification. At a classical level we expect
all of the operators to be built out of ${\rm Tr
}\left[W^2\right]$, or say $S$.

Interestingly Shore \cite{Shore:1982kh} as well as Kaymakcalan and
Schechter (KS) \cite{Kaymakcalan:1983jh} have shown that apart
from $S$ we can construct, in terms of the underlying fields, only
another independent controgradient field which transforms
covariantly under the superconformal transformations, i.e.:
\begin{eqnarray}\bar{D}^2 S^{\dag {\frac{1}{3}}}
~=~-\frac{4}{3}\frac{\bar{F}}{\bar{\varphi}^{
2/3}}-\frac{4}{9}\frac{\bar{\psi}^2}{\bar{\varphi}^{
5/3}}-\frac{4\sqrt{2}}{3}i\theta\sigma^{\mu}\partial_{\mu}\left(\frac{\bar{\psi}}{\bar{\varphi}^{
2/3}}\right)-4\theta^2 \fund \,\bar{\varphi}^{ 1/3} \ .
\end{eqnarray}
Note that the lowest component contains the $G_{\mu\nu}G^{\mu\nu}$
as well as $G_{\mu\nu}\widetilde{G}^{\mu\nu}$ operators when
expressing the fields of $S$ as in eq.~(\ref{decomp}).

We can thus define the field:
\begin{equation}\label{phi} \Phi\,=\,\bar{D}^2
S^{\dag\vspace{1cm}1/3}\,+\,r\,S^{2/3} \ ,
\end{equation}
with $r$ an unknown coefficient. $\chi$ can be naturally
introduced as:
\begin{equation}\label{chi}
\chi~=~S^{-2/3}~\Phi~=~S^{-2/3}\bar{D}^2S^{\dag 1/3}~+~r \ .
\end{equation}
This field has the right quantum numbers to describe the glueball
state we have already added in the effective theory. Note that
since $S$ acquires a non zero vacuum expectation value it is a
well defined operation to divide by powers of $S$. This relation
also tells about how the fundamental degrees of freedom are
related to $\chi$. Due to the presence of $\bar{D^2}$ acting on
$S^{\dag 1/3}$ the $G^{\mu\nu}G_{\mu \nu}$ and
$G^{\mu\nu}\widetilde{G}_{\mu \nu}$ operators are partially
contained in $\chi$. Note however that the interpolating field for
a glueball-type of state does not need to be constructed only out
of these operators.

Now let us make some considerations on the vacuum expectation
values. Using the relations (\ref{phi}-\ref{chi}) in a given super
Yang-Mills vacuum we deduce:
\begin{eqnarray}
\label{phivev}\langle \Phi\rangle~=~r~\Lambda^2\quad {\rm and}
\quad \label{chivev} \langle \chi\rangle~=~r \ .
\end{eqnarray}
This implies that the glueball field condenses, as long as $r$ is
different from $0$. We stress that glueball condensation when
supersymmetry is intact does not signal the emergence of the gluon
condensate. The latter is always guaranteed to
vanish\footnote{This can be easily checked at the effective
Lagrangian level. Since supersymmetry does not break we must have
zero vacuum energy density. However the vacuum energy density is
$\frac{1}{4}\langle \vartheta^\mu_\mu\rangle$ and
$\theta^{\mu}_{\mu}$ is the trace of the energy momentum tensor.
This is proportional to $G_{\mu\nu}^a {G}^{a ,\, \mu\nu}$, see
\cite{Sannino:2003xe}.}. An instanton computation could be able to
provide a value for $r$ directly from the microscopic theory.

The presence of the constant term $r$, related to a non zero
vacuum expectation value for $S$, is crucial. Indeed it allows us
to consider $\chi$ as an independent quantum field with respect to
$S$. We note that the classical relation (obtained by setting
$r=0$) between $\chi$ and $\bar{D}^2\,S^{\dagger 1/3}$ is the
analogous of the relation introduced by Shore \cite{Shore:1982kh}
and KS \cite{Kaymakcalan:1983jh}\footnote{Actually the KS and
Shore dimensionless field is $\chi^{-1/2}$ for $r=0$.}. However in
these works because of such a classical constraint $\chi$ was, in
practice, never an independent field. Since a controgradient field
is involved, due to the classical constraint, non holomorphic
terms for the superpotential are induced. This explains why in
\cite{Shore:1982kh,{Kaymakcalan:1983jh}} at the effective
Lagrangian level supersymmmetry was not guaranteed to remain
intact. We interpret instead the relation (\ref{chi}) as a quantum
constraint between two independent fields, $S$ and $\chi$. Now the
superpotential is holomorphic in these fields and in general
supersymmetry holds. Upgrading $\chi$ to an independent physical
field solves at once the long standing puzzle associated to the
approach used by KS and Shore. We note that the upgrade from a
classical to a quantum constraint is not a new idea, see
\cite{Intriligator:1995au} for a review.

The tree-level theory we start from is
\begin{eqnarray}
L_{\rm fund}=-\frac{2N}{3} \int d^2\theta \, (2\pi\,i \tau)\, S +
{\rm c.c.} \ ,
\end{eqnarray}
It is a standard procedure to add to this Lagrangian the one loop
corrections so that it properly accounts (in the Wilsonian scheme)
for the perturbative dynamics. Apart from the trivial factor
$-{2N}/{3}$ one gets:
\begin{eqnarray}
\int d^2\theta \,
\left[3N\,\ln\left(\frac{\Lambda}{\Lambda_0}\right)\right]S + {\rm c.c.}\ .
\end{eqnarray}
Now we want to further modify this Lagrangian to take into account
also the (non-perturbative) phenomenon of gaugino condensation. We
thus introduce the field $\chi$ and write:
\begin{eqnarray}
\label{luv}
\int d^2\theta \,\left\{
\left[3N\,\ln\left(\frac{\Lambda}{\Lambda_0}\right)\right]S
  + {\cal C}\left( S-\Lambda^3\mbox{e}^{-\frac{2\pi i k}{N}}\right) \chi+ {\rm c.c.} \right\}\ .
\end{eqnarray}
Variation with respect to $\chi$ implements gaugino condensation. To determine the constant ${\cal C}$ we note
that the coefficient of ${\rm Tr}[W^2]$, i.e. $S$, is the full
coupling constant $\hat{\tau}$.  We hence deduce:
\begin{eqnarray}
2\pi\,i \tau + 3N\ln\left(\frac{\Lambda}{\Lambda_0}\right) + {\cal
C}\,r = 2\pi\,i \hat{\tau} \ ,
\end{eqnarray}
where we have substituted the on-shell value
for $\chi$ (i.e. $r$). Here $\tau$ is the bare coupling constant. The
coefficient ${\cal C} r$ is clearly a shift of the complexified
coupling constant and as such we must have:
\begin{eqnarray}
{\cal C}r = 2 \pi\, i\, k \ .
\end{eqnarray}
This implies that the coefficient of $\chi$ in (\ref{luv}) is
$\frac{2\pi i \,{k}}{rN}~\Lambda^3 \mbox{e}^{-\frac{2\pi i k}{N}}$
(once the $N$ in front of the $S\,\ln \Lambda^3$ term is factored
out). A source term for $\chi$ that depends on $k$ (the shift of
the complexified coupling constant) has now been generated. This
is precisely what we expected for the glueball field. It is thus
tempting to identify the $\chi$ field just introduced with the
glueball superfield. This is not yet sufficient to state that we
have generated a term linear in $2\pi i \,k/N$ as a source for it.
However self-consistency with the integrating in procedure
requires the vev of $\chi$ (i.e. $r$) to have precisely the same
$k$-dependence of the vev of $S$ (see (\ref{relazione})). These
findings complete the results presented in 3.3.

Inspired by the UV identification (\ref{chi}) we might replace (\ref{luv}) with
\begin{eqnarray}
\int d^2\theta \,\left\{
\left[3N\,\ln\left(\frac{\Lambda}{\Lambda_0}\right)\right]S
  + {\cal C}\left( S-\Lambda^3\mbox{e}^{-\frac{2\pi i k}{N}}\right) \left(\, \chi  - S^{-2/3}\bar{D}^2S^{\dag
1/3}\right) + {\rm c.c.} \right\}\
\end{eqnarray}
Indeed the only difference would be a term containing a
controgradient field that, although formally written as an $F$
term, is actually a $D$ term.

\subsection{Dijkgraaf--Vafa  Connection}

A powerful tool to study the dynamics of super Yang-Mills theories
is to embed them in string theory. Within this framework, various
approaches and several line of investigations have been developed,
all going under the rather generic name of gauge-string
correspondence. One of the most interesting result which emerged
is the relation of ${\cal N}=1$ super Yang-Mills to matrix models.
This relation has been found following a long path, going through
topological string theory, superstring theory and D-branes.

However, after the matrix model structure of gauge theories has
been conjectured in this set-up \cite{Dijkgraaf:2002dh}, it has
also been possible to recover the same results in a purely field
theoretical approach \cite{Cachazo:2002ry,Cachazo:2003yc} (for a
nice and detailed review see \cite{Argurio:2003ym}).

We would like to compare our findings with these more `stringy'
approaches. According to which particular geometric model one
analyzes, there are different ways to make such a comparison (some
of them are currently under investigation). Here we see that the
inclusion of the glueball degrees of freedom in ${\cal N}=1$ super
Yang-Mills, pursued in the way we described in this paper, fits
naturally within the Dijkgraaf--Vafa (DV) approach or the related
one by Cachazo, Douglas, Seiberg and Witten, (CDSW).

Note that, even if in these approaches one deals with ${\cal N}=1$
super Yang Mills theory coupled to a chiral adjoint scalar
supermultiplet $\Phi$ with a superpotential at tree level, it is
possible to choose a quadratic superpotential, that consists just
of a mass term for $\Phi$. Assuming a large mass, it is
straightforward to integrate $\Phi$ out. In this way, at low
energy one is left with pure ${\cal N}=1$ super Yang-Mills theory.
We focus on the CDSW results (we refer the reader to
\cite{Cachazo:2002ry} for notation and further details) but one
should keep in mind that there is an obvious translation in more
geometrical terms (\`a la DV).

Within this approach it is possible to write the low energy
superpotential for the field $S$ as:
\begin{eqnarray}\label{supercdsw} W_{eff}[S]~=~2\pi
i\left(\int_AT \int_BR -\int_BT \int_AR\right) \ , \end{eqnarray}
where $A$ and $B$ are proper compact and non-compact cycles over
the complex plane, which arises as the moduli space of the adjoint
field $\Phi$. The appearance of a branch cut and non-trivial
cycles is the result of the quantum dynamics of the underlying
theory \cite{Cachazo:2002ry}. $T$ and $R$ are suitable one forms
defined over the complex plane. They have the following properties
\cite{Cachazo:2002ry,Cachazo:2002zk}: \begin{eqnarray}
\label{recipe}
\int_A T(z) dz = N&,&\hspace{1.3cm}\int_B T(z) dz = -\tau + k\ ,\\
\int_A R(z) dz = S&,&\hspace{1.3cm}\int_B R(z) dz = \frac{1}{2\pi
i}\frac{\partial {\cal F}}{\partial S} \ , \end{eqnarray} where
$N$ is the rank of the gauge group, ${\cal F}$ is called the
``prepotential'', $\tau$ is the bare complexified coupling
constant and $k$ is an arbitrary integer. We deduce the effective
superpotential:
\begin{eqnarray}\label{supermat}
 W_{eff}[S]~=~N~\frac{\partial {\cal F}}{\partial S}~+~2\pi i\tau S~-~2\pi i k S
\end{eqnarray}

In the case we are considering (the quadratic choice for the
tree-level superpotential) it is possible to compute it explicitly \cite{Cachazo:2001jy,Dijkgraaf:2002dh,Cachazo:2002ry}:
\begin{eqnarray} W_{eff}[S] =
NS\left(\log\frac{\Lambda_0^3}{S}+1\right)+2\pi i\tau S -2\pi  i k
S \ ,\end{eqnarray} where $\Lambda_0$ is an ultraviolet cut-off
arising from the integration on the non-compact cycle and $\tau$
is meant to be evaluated precisely at $\Lambda_0$. In this
paragraph we have suppressed the $2N/3$ normalization factor.

Using the knowledge of the one loop $\beta$-function, we rewrite
the superpotential as
\begin{eqnarray}\label{expdv} W_{eff}[S] =
NS\left(\log\frac{\Lambda^3}{S}+1\right) -2\pi  i k S
\end{eqnarray} where $\Lambda$ is now the dynamically generated
scale. We see that this superpotential, a part from an overall
normalization, is identical to the one in eq.~(\ref{outchi}).
Recall that the integer $k$ in (\ref{outchi}) appeared after
having integrated out the glueball superfield. In
eq.~(\ref{expdv}) instead it comes from the integration of the
meromorphic one-form $T$ along the non compact cycle. To show that
the result of the integration gives, besides the complexified
coupling constant, an integer number ($k$) is a non-trivial matter
and was proven only `on-shell' in \cite{Cachazo:2003yc}.

At this point the glueball superfield may emerge if one modifies
the prescription in eq.~(\ref{recipe}) as follows:
\begin{eqnarray}
\int_B T(z) dz = -\tau-\frac{f(\chi)}{2\pi i}\ , 
\end{eqnarray} with the requirement \begin{eqnarray}
f'(\chi_0)=0\ \mbox{and}\ f(\chi_0)=-2\pi i\,k \ . \end{eqnarray}
These are exactly the properties of the function $f(\chi)$ we
introduced in eq.~(\ref{effe}). Thus the extended VY effective
superpotential may have a natural geometric interpretation.

\subsection{The three form approach}
Long ago Gates \cite{Gates:ay} has classified all of the possible
$p$-form gauge superfields in four-dimensional space time. To the
three form is associated a real superfield $U$ (not to be confused
with the real vector gauge superfield associated to the one-form)
whose field strength is
\begin{eqnarray}
\bar{D}^2\,U \ . \label{ft}
\end{eqnarray}
The gauge transformation reads:
\begin{eqnarray}
U \rightarrow U + \frac{1}{2} \left(D^{\alpha}\Gamma_{\alpha} +
\bar{D}^{\dot{\alpha}} \bar{\Gamma}_{\dot{\alpha}} \right) \ ,
\end{eqnarray}
and it involves the gauge superfield associated to the 2-form
$\Gamma_{\alpha}$. Gates observed that the field strength in
eq.~(\ref{ft}) is equivalent to a chiral multiplet with the
pseudoscalar auxiliary field replaced by a four-from field
strength. Since the pseudoscalar auxiliary field of $S$ has a
component proportional to $G_{\mu\nu} \widetilde{G}^{\mu\nu}$
which can be rewritten as a four-form field strength one can
consider
\cite{Burgess:1995kp,Binetruy:1995hq,Farrar:1997fn,Farrar:1998rm,Cerdeno:2003us}
expressing $S$ as the field strength of $U$ via the relation:
\begin{eqnarray}
\bar{D}^2U = -\frac{S}{4} \ .
\end{eqnarray}
Due to the gauge invariance associated to the three form the
physical degrees of freedom are the ones contained in $S$. Besides
the four-form is still an auxiliary field and must be integrated
out in the end. So unless the gauge symmetry is broken or
additional chiral superfield are explicitly added no new degrees
of freedom, except for the physical ones already present in $S$,
are generated. In \cite{{Farrar:1997fn},Farrar:1998rm} and
\cite{Cerdeno:2003us} different gauge breaking terms were added to
the theory. It was also realized in \cite{Farrar:1998rm} that the
net effect of such a gauge breaking term is that now one has two
independent chiral superfields $\chi$ and $S$. $\chi$ has the same
quantum numbers we considered. We expect that a superpotential
identical to the one we found can emerge using the three-form
approach when allowing for a more general set of gauge symmetry
breaking terms while further enforcing the consistency checks.
Another possibility would be to keep gauge invariance of the three
form which describes $S$ and add a new independent field $\chi$.

\section{Conclusions}
\label{quattro}

 We have proposed an extension of
the VY effective theory which takes in to account ordinary
glueball states. The general construction principle has been to
identify first the $R$-symmetry quantum number of the chiral
superfield describing the glueball-type state. We have then used
superconformal covariance to determine the conformal weight of the
associated chiral superfield describing the glueballs. This
allowed us to construct the superpotential which still saturates
the anomalies of the underlying ${\cal N}=1$ super Yang-Mills
theory. These constraints were not sufficient to fix the
superpotential written in terms of $S$ and the glueball superfield
$\chi$.

We proposed, however, a specific form of the superpotential which
has amusing properties and passes a number of consistency checks.
{}For example we were able to integrate ``out'' and ``in'' the
field $\chi$. We obtained in the first case the standard VY theory
while in the second we deduced the extended VY effective theory.
We have shown that in the present approach the $N$ vacua of the
theory emerge due to the presence of the glueball superfield. This
fact had a natural counterpart in the geometric approach to the
effective Lagrangian theory proposed by Dijkgraaf and Vafa.
However a better understanding of this relation is needed. We have
also broken supersymmetry by adding a gluino mass. The effective
theory has led to a K\"{a}hler independent part of the
``potential'' which reproduces the glueball effective potential
for the non supersymmetric pure Yang-Mills theory.

Since the superpotential is known it will be interesting, in the
future, to investigate in some detail some physical consequences
associated, for example, to the spectrum of the theory. Moreover
the generalization of the extended VY theory to orientifold field
theories is an interesting avenue to explore. We plan to study
also super quantum chromodynamics. We will also try to gain
further insight using string theory approaches suited to describe
gauge dynamics.

\vskip 1cm
\begin{center}
\bf Acknowledgments
\end{center}  We thank  L. Bergamin, D.G. Cerde\~{n}o, P. Di Vecchia, A. Feo, G. Ferretti, J.
Louis, N. Obers, P. Salomonson, M. Shifman, A. Smilga, G. Vallone
and F. Vian for helpful discussions while P.H. Damgaard and J.
Schechter also for careful reading of the manuscript.


\begin{thebibliography}{9}


\bibitem{Intriligator:1995au}
K.~A.~Intriligator and N.~Seiberg, ``Lectures on supersymmetric
gauge theories and electric-magnetic  duality,'' Nucl.\ Phys.\
Proc.\ Suppl.\  {\bf 45BC}, 1 (1996) [arXiv:hep-th/9509066].





\bibitem{Veneziano:1982ah}
G.~Veneziano and S.~Yankielowicz, {\em An Effective Lagrangian For
The Pure N=1 Supersymmetric Yang-Mills Theory}, Phys.\ Lett.\ B
{\bf 113}, 231 (1982).






\bibitem{Sannino:2003xe}
F.~Sannino and M.~Shifman, {\em Effective Lagrangians for
orientifold theories}, arXiv:hep-th/0309252.



\bibitem{Feo:2002yi}
I.~Montvay, {\em SUSY on the lattice},  Nucl.\ Phys.\ Proc.\
Suppl.\ {\bf 63}, 108 (1998) [hep-lat/9709080];
A.~Donini, M.~Guagnelli, P.~Hernandez and A.~Vladikas, {\em
Towards N = 1 Super-Yang-Mills on the lattice}, Nucl.\ Phys.\ B
{\bf 523}, 529 (1998) [hep-lat/9710065];
For a recent review see A.~Feo, {\em Supersymmetry on the
lattice,} hep-lat/0210015.

\bibitem{Armoni:2004uu}
A.~Armoni, M.~Shifman and G.~Veneziano,
arXiv:hep-th/0403071.


\bibitem{Corrigan:1979xf}
E.~Corrigan and P.~Ramond,
Phys.\ Lett.\ B {\bf 87}, 73 (1979).



\bibitem{Shore:1982kh}
G.~M.~Shore, `Constructing Effective Actions For N=1 Supersymmetry
Theories. 1. Symmetry Principles And Ward Identities,'' Nucl.\
Phys.\ B {\bf 222}, 446 (1983).



\bibitem{Kaymakcalan:1983jh}
O.~Kaymakcalan and J.~Schechter, ``Superconformal Anomalies And
The Effective Lagrangian For Pure Supersymmetric QCD,'' Nucl.\
Phys.\ B {\bf 239}, 519 (1984).

\bibitem{Farrar:1997fn}
G.~R.~Farrar, G.~Gabadadze and M.~Schwetz, {\em On the effective
action of N = 1 supersymmetric Yang-Mills theory}, Phys.\ Rev.\ D
{\bf 58}, 015009 (1998) [arXiv:hep-th/9711166].

\bibitem{Farrar:1998rm}
G.~R.~Farrar, G.~Gabadadze and M.~Schwetz, {\em The spectrum of
softly broken N = 1 supersymmetric Yang-Mills theory} Phys.\ Rev.\
D {\bf 60}, 035002 (1999) [hep-th/9806204].

\bibitem{Gabadadze:1998bi}
G.~Gabadadze,
Nucl.\ Phys.\ B {\bf 544}, 650 (1999) [arXiv:hep-th/9808005].

\bibitem{Bergamin:2003ub}
L.~Bergamin and P.~Minkowski,
arXiv:hep-th/0301155.




\bibitem{Sannino:1997dd}
F.~Sannino and J.~Schechter, {\em  Toy model for breaking super
gauge theories at the effective Lagrangian  level}, Phys.\ Rev.\ D
{\bf 57}, 170 (1998) [hep-th/9708113];
S.~D.~Hsu, F.~Sannino and J.~Schechter, {\em Anomaly induced {QCD}
potential and quark decoupling}, Phys.\ Lett.\ B {\bf 427}, 300
(1998) [hep-th/9801097].


\bibitem{Cerdeno:2003us}
D.~G.~Cerdeno, A.~Knauf and J.~Louis, {\em A note on effective
${\cal N} = 1$ super Yang-Mills theories versus lattice results,}
hep-th/0307198.



\bibitem{Dijkgraaf:2002dh}
R.~Dijkgraaf and C.~Vafa, ``A perturbative window into
non-perturbative physics'', arXiv:hep-th/0208048.

\bibitem{Cachazo:2002ry}
F.~Cachazo, M.~R.~Douglas, N.~Seiberg and E.~Witten, ``Chiral
rings and anomalies in supersymmetric gauge theory,'' JHEP {\bf
0212} (2002) 071 [arXiv:hep-th/0211170].



\bibitem{Masiero-Veneziano}
A.~Masiero and G.~Veneziano, Nucl. Phys. {\bf B249}, 593 (1985).


\bibitem{schechter}
J.~Schechter, {\em Effective Lagrangian With Two Color Singlet
Gluon Fields}, Phys.\ Rev.\ D {\bf 21} (1980) 3393.

\bibitem{joe}
C.~Rosenzweig, J.~Schechter and G.~Trahern, Phys. Rev. {\bf D21},
3388 (1980); P.~Di Vecchia and G.~Veneziano, Nucl. Phys. {\bf
B171}, 253 (1980); E.~Witten, Ann. of Phys. {\bf 128}, 363 (1980);
P.~Nath and A.~Arnowitt, Phys. Rev. {\bf D23}, 473 (1981);
A.~Aurilia, Y.~Takahashi and D.~Townsend, Phys. Lett. {\bf 95B},
65 (1980); K.~Kawarabayashi and N.~Ohta, Nucl. Phys. {\bf B175},
477 (1980).

\bibitem{MS}
A.~A.~Migdal and M.~A.~Shifman,
Phys.\ Lett.\ B {\bf 114}, 445 (1982);
J.~M.~Cornwall and A.~Soni,
Phys.\ Rev.\ D {\bf 29}, 1424 (1984);
Phys.\ Rev.\ D {\bf 32}, 764 (1985).

\bibitem{SST}
A.~Salomone, J.~Schechter and T.~Tudron, Phys. Rev. {\bf D23},
1143 (1981); J.~Ellis and J. Lanik, Phys. Lett. {\bf 150B}, 289
(1985); H.~Gomm and J.~Schechter, Phys. Lett. {\bf 158B}, 449
(1985); F.~Sannino and J.~Schechter,
Phys.\ Rev.\ D {\bf 60}, 056004 (1999) [hep-ph/9903359].


\bibitem{Kovner-Shifman}
A.~Kovner and M.~A.~Shifman,  Phys.\ Rev.\ D {\bf 56}, 2396 (1997)
[hep-th/9702174].



\bibitem{EHS}
N.~Evans, S.D.H.~Hsu and M.~Schwetz, Phys. Lett. B {\bf 404}, 77
(1997); Nucl. Phys. {\bf B484}, 124 (1997); N.~Evans, S.D.H.~Hsu,
M.~Schwetz, S.B.~Selipsky, Nucl. Phys. {\bf B456}, 205 (1995).


\bibitem{Hisano:1997ua}
J.~Hisano and M.~A.~Shifman, ``Exact results for soft
supersymmetry breaking parameters in  supersymmetric gauge
theories,'' Phys.\ Rev.\ D {\bf 56}, 5475 (1997) [hep-ph/9705417].

\bibitem{Cachazo:2002zk}
F.~Cachazo, N.~Seiberg and E.~Witten,
``Phases of N = 1 supersymmetric gauge theories and matrices,''
JHEP {\bf 0302}, 042 (2003)
[arXiv:hep-th/0301006].

\bibitem{Cachazo:2001jy}
F.~Cachazo, K.~A.~Intriligator and C.~Vafa,
``A large N duality via a geometric transition,''
Nucl.\ Phys.\ B {\bf 603}, 3 (2001)
[arXiv:hep-th/0103067].


\bibitem{Cachazo:2003yc}
F.~Cachazo, N.~Seiberg and E.~Witten, ``Chiral Rings and Phases of
Supersymmetric Gauge Theories,'' JHEP {\bf 0304} (2003) 018
[arXiv:hep-th/0303207].

\bibitem{Argurio:2003ym}
R.~Argurio, G.~Ferretti and R.~Heise, ``An introduction to
supersymmetric gauge theories and matrix models,''
arXiv:hep-th/0311066.






\bibitem{Gates:ay}
S.~J.~J.~Gates, ``Super P Form Gauge Superfields,'' Nucl.\ Phys.\
B {\bf 184}, 381 (1981).



\bibitem{Burgess:1995kp}
C.~P.~Burgess, J.~P.~Derendinger, F.~Quevedo and M.~Quiros, {\em
Gaugino condensates and chiral linear duality: An Effective
Lagrangian analysis}, Phys.\ Lett.\ B {\bf 348}, 428 (1995)
[arXiv:hep-th/9501065].

\bibitem{Binetruy:1995hq}
P.~Binetruy, M.~K.~Gaillard and T.~R.~Taylor, {\em ``Dynamical
supersymmetric breaking and the linear multiplet},'' Nucl.\ Phys.\
B {\bf 455}, 97 (1995) [arXiv:hep-th/9504143].






\end{thebibliography}
\end{document}